\documentclass[
    twocolumn,
    aps,
    prl,
    superscriptaddress,
    longbibliography,
]{revtex4-2}

\usepackage{graphicx,color,xcolor}
\usepackage[
    colorlinks=true,
    urlcolor=blue,
    linkcolor=blue,
    citecolor=blue,
    breaklinks,
]{hyperref}

\usepackage{amsmath}
\usepackage{amsfonts}
\usepackage{amssymb}
\usepackage{braket}
\usepackage{mathtools}
\usepackage{makecell} 

\begin{document}

\title{The universality class of the first levels in low-dimensional 
 gravity}

\author{Alexander Altland}
    \email{alexal@thp.uni-koeln.de}
    \affiliation{Institute for Theoretical Physics, Zülpicher Str. 77a, 50937, Cologne, Germany}

\author{Jeremy van der Heijden}
    \email{jeremy.vanderheijden@ubc.ca}
    \affiliation{Department of Physics and Astronomy, University of British Columbia, 6224 Agricultural Road, Vancouver, B.C. V6T 1Z1, Canada}

\author{Tobias Micklitz}
    \affiliation{Centro Brasileiro de Pesquisas F\'isicas, Rua Xavier Sigaud 150, 22290-180, Rio de Janeiro, Brazil} 

\author{Moshe Rozali}
    \email{rozali@phas.ubc.ca}
    \affiliation{Department of Physics and Astronomy, University of British Columbia, 6224 Agricultural Road, Vancouver, B.C. V6T 1Z1, Canada}

\author{Joaquim Telles de Miranda}
    \email{joaquim@cbpf.br}
    \affiliation{Centro Brasileiro de Pesquisas F\'isicas, Rua Xavier Sigaud 150, 22290-180, Rio de Janeiro, Brazil}

\date{\today} 

\begin{abstract}
We investigate the physics of a small group of quantum states defined above the
sharply defined ground state of a chaotic ensemble.  This  `universality class
of the first levels' (UFL) is realized in the majority of `synthetic' random
matrix models but, for all we know, in only one microscopically defined system:
low-dimensional gravity. We discuss the physical properties of these states,
notably their exceptional rigidity against external perturbations, as quantified
by the so-called quantum state fidelity. Examining these structures through the
lenses of random matrix  and string theory, we highlight their relevance to the
physics of low-dimensional holographic principles.

\end{abstract}

\maketitle

{\it Introduction:---} The advent of the SYK
model~\cite{rosenhausIntroductionSYKModel2019} has made systems exhibiting
`maximal chaos' all the way down to their ground states  a focus of attention.
For generic representatives of this class, changes in comparatively few system
parameters (think of the interaction constants of a many-body Hamiltonian)
affect exponentially many levels, turning the spectrum, and in particular the
ground state itself into broadly distributed
objects~\cite{Garcia-Garcia2016,berkoozFullSolutionLarge2019}. The situation is
different for systems that are `dense', in the sense that the number of
statistically independent parameters is comparable to their Hilbert space
dimension, $D$~\cite{Altland:2024ubs}. For these systems,  the spectral edge and
the first few states above  define a miniature `universality class of the first
levels' (UFL for brevity, throughout), with physical properties not realized
elsewhere in the spectrum: both the position of the UFL levels and, as we are
going to show in this Letter, their wave functions display exceptional rigidity
under configurational averaging.  Intuitively, this behavior results from the
squeezing of mutually repelling levels between a non-fluctuating edge and the
bulk of the spectrum. Representing fine-grained system properties at scales of
the many-body level spacing, these are non-perturbative structures, which cannot
be captured by semiclassical $1/D$ expansion schemes.

\begin{figure}
    \centering
        \includegraphics[width=0.8\linewidth]{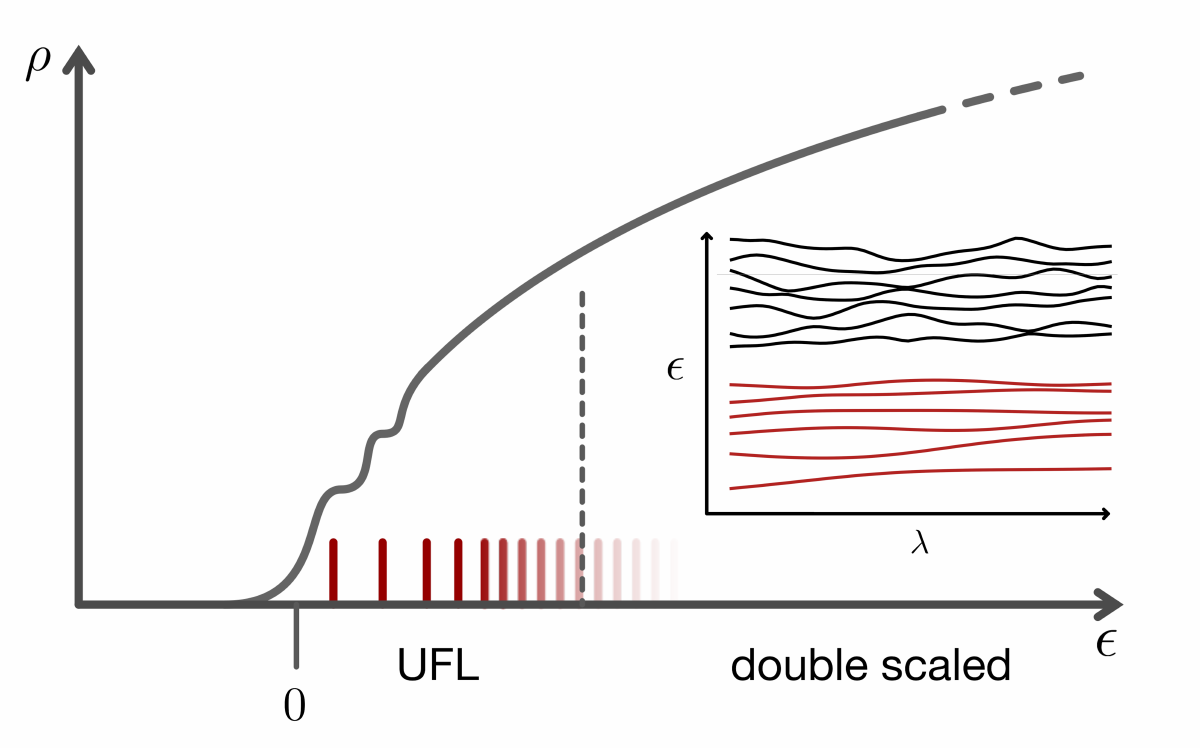} 
        \caption{Schematic structure of the near edge spectrum of a dense
        system. Close to the edge, the level spacing increases and statistical
        fluctuation of levels freeze out, as witnessed by the Airy-oscillatory
        average spectral density. Individual realizations of sparse systems show
        a similar edge structure, however, large fluctuations of the edge
        \emph{position}, indicated as $\epsilon=0$, remove these signatures in
        average quantities and correlation functions. Inset: Energy-levels as a
        function of parametric  perturbations demonstrating the rigidity of near
        edge states (bottom) compared to generic bulks (top) states.}
        \label{fig:UFLSchematic}
    \end{figure}

Dense systems harboring a UFL are mostly of synthetic nature, examples including
 random matrix ensembles~\cite{GuhrRMT1998}, quantum
 graphs~\cite{gnutzmannQuantumGraphsApplications2006}, or quantum circuits
 subject to `dense' Haar distributed randomness~\cite{FisherRandomCircuit}.
 However, a prime realization of a naturally occurring dense system is
 low-dimensional gravity. Considered as a sum over  geometries, lattice
 discretizations~\cite{ashtekarShortReviewLoop2021} define  structures with
 about as many ensemble parameters as degrees of freedom, which is the dense
 scenario. 

For the paradigmatic example of two-dimensional
Jackiw-Teitelboim (JT) gravity this view has been made concrete by demonstrating
quantitative equivalence between the expansion of the gravitational path
integral over geometries of ascending topological complexity and the $1/D$
expansion of a random matrix ensemble \cite{Saad:2019lba}. Perturbative in
$1/D$, these probes are not yet fine-grained enough to witness the presence of
the UFL. However, it has turned out that an extension of the semiclassical
gravitational path integral to topological string theory provides the required
precision \cite{Post:2022dfi, Altland:2022xqx}.

While previous analyses of chaos in the gravitational context have focused on
spectral distributions, recent work \cite{Berry2020,Pandey2020} has pointed out
that the \emph{geometry of quantum states} includes more sensitive information.
Probes such as the quantum geometric tensor
\cite{Provost:1980nc,PhysRevLett.99.095701}, the adiabatic gauge potential
\cite{Kolodrubetz2017,Berry2009}, or fidelity susceptibilities
\cite{Sels2021,Sierant2019} describe how typical wave functions  deform under
the influence of external parameter variations, providing highly resolved
portraits of specific chaotic quantum systems.

In this Letter, we apply this framework to a characterization of UFL quantum
states. Focusing on the distribution of fidelity susceptibilities, we will
demonstrate resilience to perturbations parametrically exceeding that of generic
bulk states. While in the bulk ensemble variations allow only for statistical
characterization averaged over many level spacings, these findings apply to an
$\mathcal{O}(1)$ group of almost completely pinned states. In this sense, our
present analysis may be the first description of black hole quantum mechanics
tuned to the precision of individual states  (if in the toy context of
two-dimensional gravity). We will also discuss the implications of our findings
for the construction of holographic boundary shadows of two-dimensional gravity.

{\it Signatures of the spectral edge:---}The physics of the UFL is intimately
linked to the interpretation of the spectral edge as a symmetry breaking phase
transition, with the energy from the edge, $\epsilon$, as the control parameter, and the average
spectral density $\langle \rho(\epsilon) \rangle$ an order parameter. The broken
symmetry is causality, i.e.  the difference between retarded and advanced
resolvents, $G^\pm(\epsilon)\equiv (\epsilon \pm i0  - H)^{-1}$,   which outside
the edge at $\epsilon\equiv 0$ vanishes on average $ \langle
{\rm tr}(G^-(\epsilon)-G^+(\epsilon))\rangle=0$ (up to corrections exponentially
small in Hilbert space dimension), while inside it is proportional to the
spectral density order parameter. 

The universal `$\phi^4$-theory' of this phase transition is the 
Kontsevich  model \cite{Kontsevich:1992ti}, an integral over low-dimensional `flavor'-matrices, $A$,  
with action 
\begin{align}
    \label{eq:Kontsevich}
    Z(X)\equiv \int dA \,e^{S(A)},\quad  S(A)=c\, {\rm str}\left( XA +\frac{1}{3}A^3\right),
\end{align}
where, $X =
\textrm{diag}(\epsilon_1,\dots,\epsilon_m|\epsilon_1,\dots,\epsilon_n)$ is  a
diagonal graded matrix containing $m$ (`bosonic') and $n$ (`fermionic') energies
at which the system is probed, $A$ are supermatrices (matrices containing
Grassmann and commuting variables~\cite{Efetbook}) of dimension $m+n$, and the
integration is over a flat measure. The symbol `${\rm str}$' denotes the supermatrix
trace operation, and $c$ a coupling constant.

The Kontsevich model offers complementary perspectives: In quantum chaos, it
defines an extension of the  nonlinear $\sigma$-model~\cite{Efetbook} from the
bulk spectrum to the edge~\cite{SonnerAltland21}, providing a description of
both  energy levels and wave functions. For example, for $m,n=1$, with just one
probe energy $\varepsilon_1=\epsilon_1=\epsilon$, a straightforward stationary
phase analysis yields the estimate $\langle\rho(\epsilon)\rangle\simeq (c/\pi)\epsilon^{1/2}$,
defining a semiclassical approximation for the  spectral density. By contrast,
the exact integral over $2$-dimensional  
supermatrices yields $\langle \rho(\epsilon) \rangle=c^{2/3}\left(\tilde
\epsilon \textrm{Ai}^2(-\tilde \epsilon)+(\textrm{Ai}'(-\tilde \epsilon))^2
\right) $, $\tilde \epsilon=\epsilon c^{2/3}$, where the appearance of
$\textrm{Ai}$ is explained by the resemblance of $\int_A \exp(S(A))$ to integral
representations of the Airy function~\cite{Vallee2010}. Physically, the refined
result characterizes the spectral edge with oscillatory near-edge spectral
density, qualitatively indicated in Fig.~\ref{fig:UFLSchematic}, where
oscillation maxima mark the rigid positions of UFL energy levels, and minima
describe their repulsion. Their statistical distribution, witnessed by the
rounded profile of the average spectral density, is probed by correlation
functions, such as $\langle \rho(\epsilon_1)\rho(\epsilon_2) \rangle$, which can
be computed with a little more effort from a $m,n=2$  
 version of the model~\cite{Altland:2024ubs}.  

Within the gravitational context, the perturbative expansion of
the Kontsevich quantitatively matches the topological expansion of the gravitational
path integral~\cite{Saad2024,Blommaert2023}. Beyond perturbation theory it features as the effective
theory describing 
 correlations in topological string
theory~\cite{Aganagic:2003qj, Altland:2022xqx} at the level required to describe,
and physically interpret the UFL. In the following, we apply this framework to
describe the rigidity of ground state wave functions to parameter variations in
terms of the so-called quantum geometric tensor (QGT). We will first compute 
the distribution of diagonal elements of the QGT
 employing a variant of Eq.~\eqref{eq:Kontsevich}, then interpret our
findings in terms of string fluctuation degrees of freedom, and finally comment
on implications for the holographic correspondence.

{\it State geometry and fidelity susceptibility:---}Quantum state geometry, i.e.
the metrics according to which individual states $\ket{n}$ of a system deform
under parameter variations, is encoded in an object known as  the quantum
geometric tensor, $g^{(n)}=g_{\alpha \beta}d\lambda^\alpha d\lambda^\beta$.  For
a Hamiltonian perturbed as $H\to H + \sum_\alpha \lambda_\alpha H_\alpha$, with
eigenstates $\ket{n}\to \ket{n_\lambda}$, its elements are defined as
\begin{align}
\label{eq:QGT_def}
g_{\alpha\beta}^{(n)} 
&= 
    \langle \partial_\alpha n | \partial_\beta n \rangle 
    - 
    \langle \partial_\alpha n | n \rangle \langle n | \partial_\beta n \rangle
.
\end{align}
where  $|\partial_\alpha
n\rangle=\partial_{\lambda_\alpha}|n_\lambda\rangle|_{\lambda=0}$. We here focus
on the diagonal elements $g_{\alpha\alpha}$, the \emph{fidelity
susceptibilities} (FS), as the most
direct measure for the response of states to perturbations. Straightforward
first order wave function perturbation theory yields the alternative
representation
\begin{align}
    \label{eq:FSSpectral}
    g^{(n)}_{\alpha\alpha}=\sum_{m\not=n}
\frac{|\langle n|H_\alpha|m\rangle|^2}{(E_n-E_m)^2}~,
\end{align}
highlighting the combined influence of spectral (denominator) and wave-function
(numerator) correlations on this observable, and energies $E_n$ are measured with respect to the band center. Reflecting the stochastic nature of both, the FSs are
distributed variables, and we aim to compute their distribution 
\begin{align} \label{eq:FS_distr_def}
P_E(g) 
&= 
\sum_n 
\langle 
\delta(E - E_n)\delta(g-g_{\alpha\alpha}^{(n)})
\rangle~,
\end{align} 
for the quantum states of a dense chaotic system with its lower edge at $E=-E_0$ in response to a
generic deformation $\lambda_\alpha$. Here,   
 $\langle ...\rangle$ denotes ensemble averaging over a Gaussian distributed
 perturbations $H_\alpha$, and over a distribution of $H$, where
 it will be interesting to compare three different settings: (i) diagonal
 matrices with randomly distributed matrix elements, (ii)
 generic states $E$ inside the spectrum of invariantly distributed $H$'s, i.e. 
 distributions satisfying $P(H)dH=P(UHU^\dagger) dH$, and (iii) the same, but for $E$ level spacing close to
 the edge, the UFL. These cases represent opposite extremes in that  (i)  has neither spectral nor state correlations, representing an
 ensemble of `integrable systems', while (ii,iii) are maximally correlated/chaotic.

\begin{figure}[tt!]
    \centering
    \vspace{-.6cm}
    \hspace{10pt}
    \includegraphics[width=\linewidth]{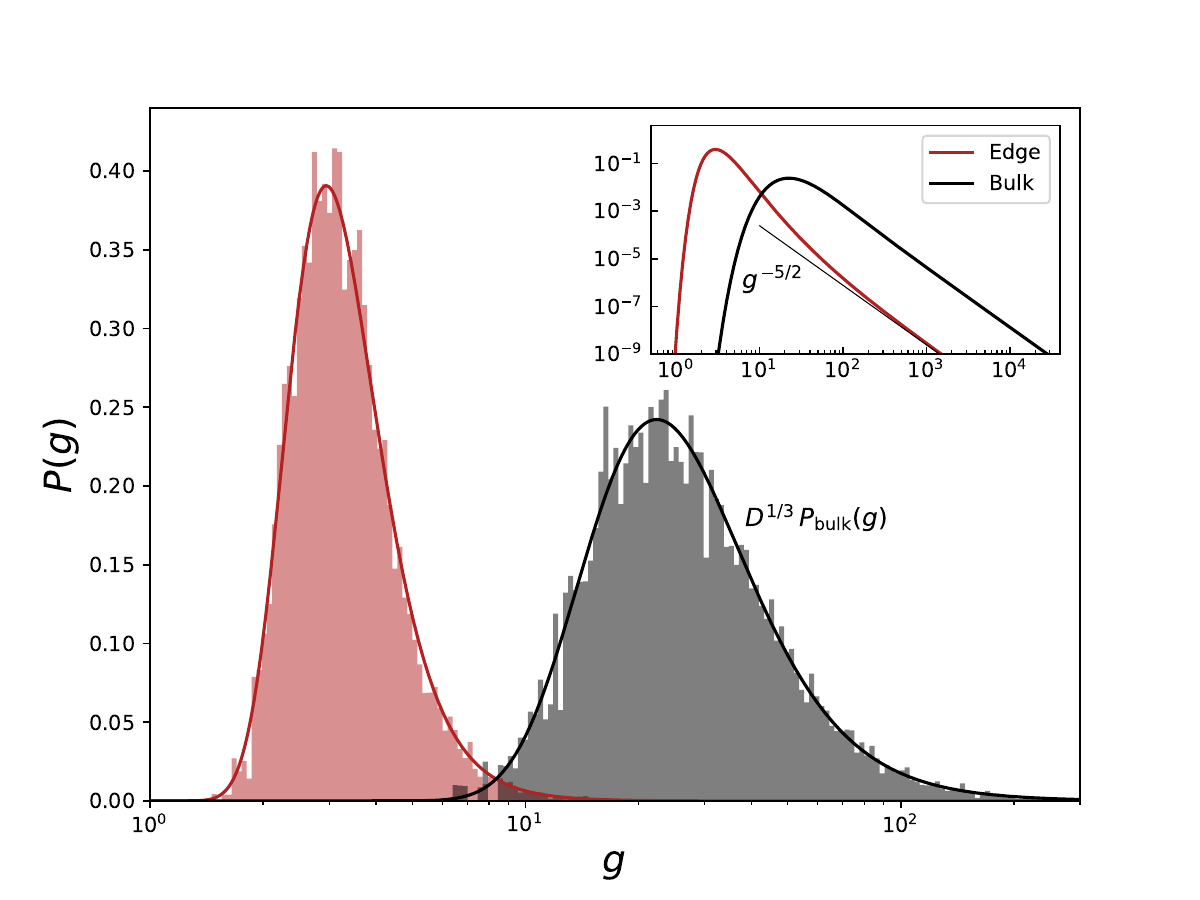}
    \caption{FS-distributions for near-edge (red)
    and bulk (black) states are compared to histograms generated from an ensemble of 
    $5000$ matrices of dimension $D=1000$. (Bulk distribution scaled by 
    $D^{1/3}$ for better visibility).
    Inset: FS-distributions for near-edge and bulk states, 
    with asymptotic scaling $g^{-5/2}$ added for reference.
    }
    \label{fig:fig2}
    \end{figure}

For each of these settings, a sequence of nontrivial but standard
operations~\cite{Penner:2020cxk,OppenPhysRevLett.73.798,supp_mat} leads to the integral representation 
$P_E(g) \propto 
\int_{-\infty}^{\infty} dz\, z Z(z)  e^{-Dgz^2}$, where  `$\propto$'
means equality up to  normalization, and the integration is over a $z$-contour
infinitesimally shifted into the
upper complex plane. The kernel  
$Z(z)=\lim_{x_i\to 0}Z(X)$,  
with 
\begin{align}
\label{eq:det_ratio}
Z(X) 
&= 
\left\langle \frac{
\prod_{i = 1}^{2+\beta}
\det( H - E + x_i)}{
\prod_{s=\pm}\det\left( H - E + sz \right)
}  
\right\rangle~, 
\end{align}
effectively defines a probability generating function with `source matrix' $X =
\textrm{diag}(z,-z|x_1,\dots,x_n)$. Here $\beta=0$ for case (i), while the
presence of two extra determinants, $\beta=2$, for (ii,iii) reflects the higher
level of correlations in these cases. For (i), the averaging of  $H$'s diagonal
elements over a box distribution and energy values $E$ 
in the center of the band is straightforward, and so is the subsequent Gaussian
$z$-integral. As a result, one obtains $P(g)\propto g^{-3/2}e^{-D/g}$, a
so-called Levy stable distribution characterized by fat tails at large $g$, and
the non-existence of statistical moments. This tendency towards large
susceptibilities is a consequence of the statistical independence of closeby
energy values in the model (i) of uncorrelated energy levels.

\begin{figure}[tt!]
\centering
    \hspace{10pt}
    \includegraphics[width=.78\linewidth]{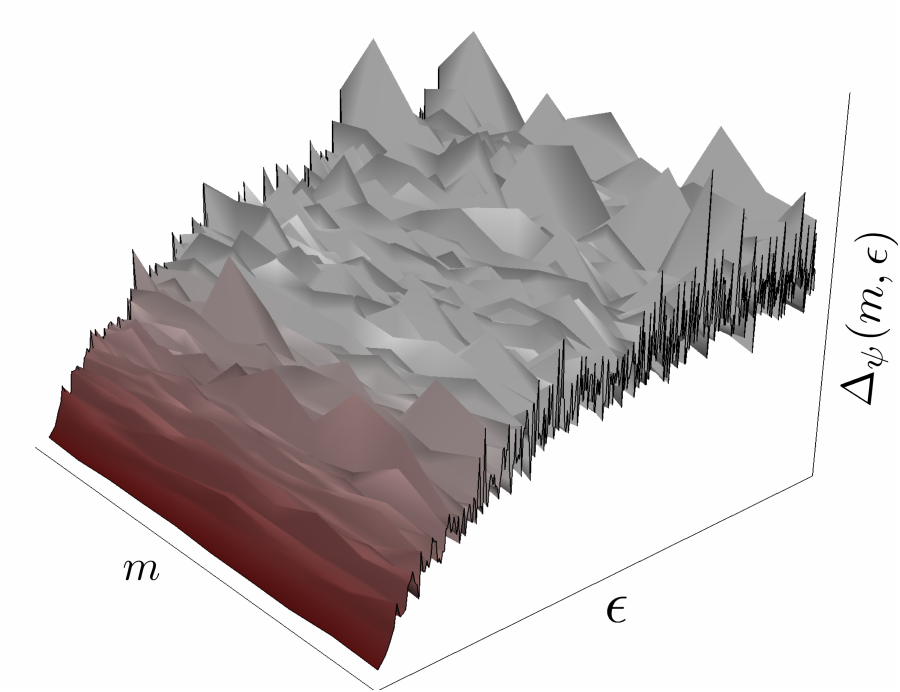}
    \vspace{.2cm}
    \caption{
    Moduli of wavefunction differences 
    $\Delta_\psi(m,\epsilon)
    \equiv ||\psi_m(0, \epsilon)|-|\psi_m(\lambda, \epsilon)||$, 
    for a fixed small perturbation $\lambda = 10^{-7}$  
    as a function of wave-component $m$ and energy $\epsilon$, 
    here for a 
    system of dimension $D=1000$, 
    averaged over $100$ samples. }
    \label{fig:fig3}
    \end{figure}

{\it Supersymmetry:---}To compute the fidelity distribution of the correlated
models (ii,iii), we employ an adapted variant of the Kontsevich model. 
Referring for details to the supplemental material, we here highlight cornerstones of the
computation which afford a gravitational perspective to be discussed below.  Our
starting point is a representation of the determinants in
Eq.~\eqref{eq:det_ratio} as Gaussian integrals over complex commuting
(denominator) and Grassmann (numerator) variables respectively. The subsequent
average over the distribution of $H$ generates correlations between these
variables, which can be decoupled by one of several available variants of
Hubbard-Stratonovich transformations \cite{Guhr2006ArbitraryUI,Guhr2009} in
terms of a 6-dimensional supermatrix, $A$. 
In the immediate vicinity of the
edge, the supermatrix integral reduces to Eq.~\eqref{eq:Kontsevich}, with
$c=D$ the Hilbert space dimension and $X=(z,-z|x_1,x_2,x_3,x_4)$. 

Turning to the computation of that integral, its structure suggests
working in polar coordinates, $A=TWT^{-1}$,  
with radial degrees of freedom contained in $W\equiv
\mathrm{diag}(u_1,u_2|-w_1,-w_2,-w_3,-w_4)$ and angular matrices $T$, which can
be integrated out via a super-variant of the Itzykson-Zuber (IZ) integration
identity \cite{superitzyksonzuber}. As a result we obtain the intermediate
representation
\begin{align}
\label{eq:Z_radial_integrals}
Z(X) \propto
\int DW\, \frac{{\rm s}\Delta(W)}{{\rm s}\Delta(X)} e^{S(W)}~,
\end{align}
where  at the edge the action reduces to the Kontsevich action \eqref{eq:Kontsevich} evaluated on the
radial matrix $A=W$, $DW$ 
 is the flat measure of radial coordinates, and 
${\rm s}\Delta$ the super-Vandermonde
determinant~\footnote{The super-Vandermonde determinant of a diagonal matrix
$Y=\mathrm{diag}(v_1,v_2|y_1,y_2,y_3,y_4)$ is defined as ${\rm s}\Delta(Y)
\equiv \frac{(v_2-v_1)\prod_{j<i} (y_i -
y_j)}{\prod_{i=1}^4\prod_{j=1}^2(y_i - v_j)}$.}. 

The final integral over radial variables now depends on whether we are probing
states inside the spectrum, 
$ E + E_0
\equiv
\epsilon 
\gg \Delta_\textrm{e}\sim D^{-2/3}$,
where $\Delta_\textrm{e}$ is the near edge level spacing, model (ii), or values
inside the domain of the UFL, $\epsilon =\mathcal{O}(\Delta_\textrm{e})$, model
(iii). 
In the former case, the integral  can be evaluated by stationary phase
methods, leading to the `correlated Levy distribution'~\cite{Penner:2020cxk,Sierant2019}
\begin{align}
  P(g) 
  &\propto  
  p\left(gD/\rho_\epsilon^2\right) e^{-\rho_\epsilon^2/Dg}, \nonumber \\ 
  p(x) &= \frac{3}{4}x^{-5/2} + x^{-7/2} + x^{-9/2}, 
\end{align}
with $\rho_\epsilon \equiv \langle \rho(\epsilon)\rangle \propto D \epsilon^{1/2}$. Compared to the model (i), the stronger attenuation at large
$g$-values, $P(g)\sim g^{-5/2}$, $g\gg D$, reflects the level repulsion of
chaotic systems, which makes anomalously large contributions to the r.h.s. of
Eq.~\eqref{eq:FSSpectral} less likely~\cite{Berry2020,Pandey2020}.

This principle finds a yet more radical manifestation as we turn to the edge, model
(iii). While saddle point analysis no longer is an option, the integral
Eq.~\eqref{eq:Z_radial_integrals} turns out exactly doable and yields $Z(z)$ 
in terms of Airy functions~\cite{supp_mat}. 
Gaussian integration then yields our main result
\begin{align}
  \label{eq:result_edge}
  P(g) 
  &\propto  
  p\left(\frac{g}{D^{1/3}}\right) e^{-\frac{D}{12 g^3}}, \nonumber \\
  p(x)&=c_5 x^{-5/2}+c_7 x^{-7/2}+ \dots + c_{17}x^{-17/2},
\end{align}
and numerical coefficients of $\mathcal{O}(1)$~\cite{supp_mat}. 
This is again a heavy tailed distribution, where the dominant power law
$x^{-5/2}$ is identical to that of model (ii), indicating the same high level
of correlation between neighboring states. Crucially, however, this power law is defined
with reference to the scaling variable $x=g D^{-1/3}$, implying that even large values
$x\gtrsim 1$ correspond to parametrically smaller values of the susceptibility
than in the bulk. This higher level of state inertia reflects the rigid spacing
of a sparse collection of states at the edge. Conversely, the dependence of the
exponent on the same scaling combination prevents anomalously \emph{small}
values $g\lesssim D^{1/3}$, leading to a distribution parametrically narrower
than in the bulk. 

Fig.~\ref{fig:fig2} shows that these results are in excellent
agreement with numerical simulations~\footnote{We here consider edge eigenstates with energies in the range $\pm
\Delta_\textrm{e}/10$. A constant shift
$g \to g + 0.47$ has been added to Eq.~\eqref{eq:result_edge} to compensate
finite size effects, and negligible in the limit 
$D\to\infty$. For bulk states we fit the energy to $\epsilon \approx 9.5\, \Delta_\textrm{e}$.}. Fig.~\ref{fig:fig3} illustrates how  wave functions
described by this statistics maximize entropy in Hilbert space (the $m$-axis) in
terms of Gaussian distributions whose sensitivity  to perturbations
is  reduced for the few terminal states forming the UFL. 

\begin{figure}[tt!]
        \centering
        \includegraphics[width=0.6\linewidth]{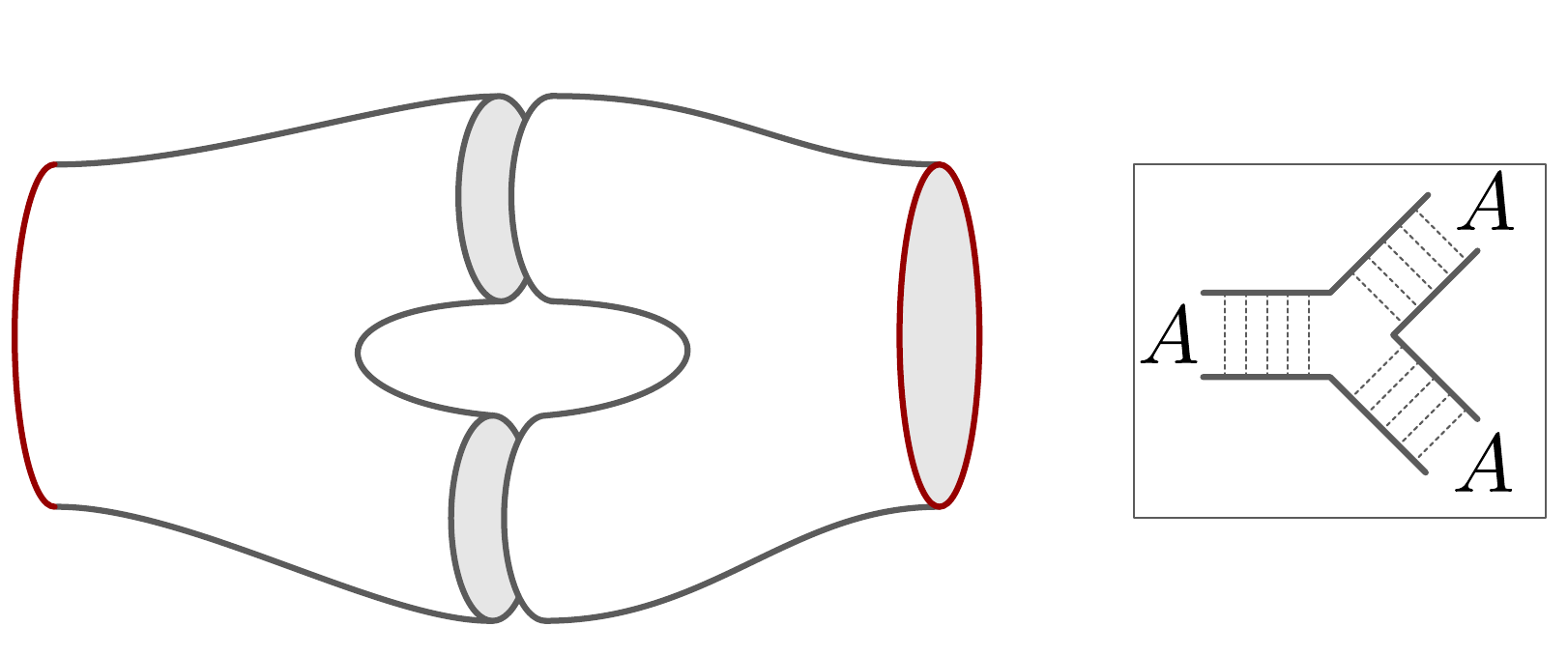}
         \caption{Cartoon illustrating the generation and recombination of `mini-universes' in the gravitational context, and its representation through the cubic vertices of the matrix theory Eq.~\eqref{eq:Kontsevich}.}
        \label{fig:fig4}
        \end{figure}

{\it String theory:---} We conclude by taking a look at the fidelity
distribution from a gravitational perspective. As mentioned above, JT gravity is
an integral over fluctuating surface geometries, see Fig.~\ref{fig:fig4} for a
cartoon representation of a surface with a single handle, and two boundaries
with circle topology, as required for the computation of spectral two-point
correlation functions \cite{Saad:2019lba}. Reading these structures from left to
right, they describe the generation and recombination of circular
`mini-universes' via cubic scattering amplitudes, as indicated by the cut
structure. These processes correspond to the likewise cubic matrix scattering
amplitudes described by the Kontsevich model \cite{Altland:2020ccq}, a
connection made quantitative through so-called ribbon-diagram
constructions~\cite{Altland:2024ubs}.  However, they
 fall short of representing quantum states of the UFL. First, the
ratios of determinants featuring in Eq.~\eqref{eq:det_ratio} do not afford a
representation in terms of surfaces with finite number of boundary insertions.
Second, an expansion in terms of surface topologies, equivalent to a
perturbative summation of matrix diagrams, does not resolve individual boundary
states. 

Both problems are elegantly solved by an effective string theory, possessing
JT gravity as its semi-classical limit \cite{Post:2022dfi}. This framework, portrayed in more detail
in the end matter, describes two-dimensional gravity as a conformal field theory
of a scalar field $\Phi$, obtained by dimensional reduction from a six-dimensional
parent geometry. The integration over open string excitations involved in this
reduction introduces the equivalent of an ensemble average, and splitting of
closed string excitations describes the mini-universe generation mentioned
above. Crucially for us, the representation of the observable
Eq.~\eqref{eq:det_ratio}, and its non-perturbative analysis are concrete options
within this framework. As detailed in the end matter, the intermediate steps of
this analysis quantitatively reproduce Eq.~\eqref{eq:Z_radial_integrals},
providing its super-mathematical structures with a physical
interpretation in terms of (anti)-brane correlations. 
The main conclusion of this analysis is that the string
theoretical framework provides access to the quantum states of the UFL in terms
of correlation functions which do not afford an interpretation in terms of
semiclassical geometry fluctuations, but one in terms of string correlations.

{\it Discussion:---}The above construction shows the quantitative equivalence
of three cornerstones in the theory  of dense chaotic systems:
the quantum state distribution Eq.~\eqref{eq:det_ratio} evaluated for ensembles of
random matrix Hamiltonians, the same observable in a theory of
topological strings, and the equivalence of both to the Kontsevich matrix model
Eq.~\eqref{eq:Kontsevich}. In this way, we have  found that the universality
class of the first levels above the edge is described by
a maximum entropy distribution distinguished by exceptional resilience  to perturbations.

The emergence of that universality class in two-dimensional holography is owed
to the confluence of two individually unique principles. First, the presence of
a non-fluctuating edge, and second the fact that JT-incorporates its own
ensemble. In two-dimensional gravity, the latter is  realized through
the integration over `hidden' open string degrees of freedom integrated out upon
dimensional reduction of JT's parent KS theory. These structures are not present in naturally
defined quantum boundary theories --- such as ensembles of SYK models ---, where `natural'
means theories defined in terms of a few-body chaotic Hamiltonians. Referring
back to the beginning of the Letter, these are \emph{sparse} theories where the
averaging over polynomially few parameters in an exponentially high dimensional
Hilbert space leads to large edge fluctuations eradicating all the structures
discussed above. In this sense, the fine-grained string theoretical probe
reveals a weak link in the low-dimensional holographic principle which
requires further investigation. 

\paragraph*{Acknowledgment:---}T.~M.~acknowledges financial support by Brazilian agencies CNPq and FAPERJ, and J.~T.~M.~
financial support by Brazilian agency CAPES. J.~v.~d.~H.~acknowledges support from the National Science and Engineering Research Council of Canada (NSERC) and the Simons Foundation via a Simons Investigator Award. Work supported by CRC183 of the Deutsche Forschungsgemeinschaft (DFG) Project No. 277101999, CRC 183
(project A03). The work is supported by a Discovery grant from NSERC.
{\bf Data and materials availability:} Processed data and python script used to generate 
Fig.~\ref{fig:fig2} and Fig.~\ref{fig:fig3} are available in Zenodo with identifier 10.5281/zenodo.15508578 \cite{Zenodo2025}.

\newpage 
\appendix

\section{Gravity perspective of the UFL}
\label{sec:GravityPerspective}

We here discuss how the UFL presents itself from the perspective of gravity
and how the latter connects to the concepts discussed in the main
text. The theory of gravitation relevant to us is called Jackiw-Teitelboim (JT) gravity, two-dimensional Einstein gravity
coupled to a scalar (`dilaton') field. This theory famously describes the
perturbative contents of a matrix model with near edge spectral density
$\tilde \rho(\epsilon)=\frac{1}{4\pi^2}\sinh 2\pi \sqrt{\epsilon}\sim \sqrt{\epsilon}$ exhibiting the
characteristic non-analyticity of a dense model. (Here $\tilde \rho(\epsilon)$ is normalized to one.) The correspondence is
established by comparing the expansion of the JT path integral in its coupling
constant $\exp(S_0)$, where $S_0\sim 1/G_N$ is the gravitational constant, ---
essentially an expansion in geometries of ascending topological complexity aka
mini-universe expansion --- to the $D^{-1}$-expansion of a matrix theory with
dimension $D\sim \exp(S_0)$~\cite{Saad:2019lba}. Intriguingly, JT gravity
describes ensemble averaged properties of the matrix theory, while individual
members of the ensemble do not have a gravitational counterpart at this level. 

As discussed in the main text, the UFL is perturbatively invisible, and hence
does not have a gravitational representation within JT theory. However, there
exists an extended framework, known as Kodaira-Spencer (KS)
theory~\cite{Bershadsky:1993cx}, which reproduces the topological expansion of
the JT path integral in perturbation theory and beautifully includes the UFL
beyond. KS theory is obtained from a theory of topological strings propagating
in a six-dimensional non-compact Calabi-Yau (CY) manifold defined as the
solution set of the complex equation 
\begin{equation} \label{eq:CY}
    H(x,y)-uv =0~,
\end{equation}
where our present discussion applies to the choice $H(x,y)\approx y^2-x$. The
two-dimensional reduction of KS theory~\cite{Dijkgraaf:2007sx} relevant to our
comparison to JT gravity and matrix theory is obtained by  integrating out
string degrees of freedom. It results in  an effective theory of a field
$\Phi(x)$ living on the two-dimensional \emph{spectral curve}, $\Sigma$, defined
as the solution set of the equation $H(x,y)=0 \Rightarrow y=y(x)$. Technically,
$\Phi$ is a chiral boson, governed by an action containing a cubic interaction. 

All elements of this framework, i.e. the boson field operators, its action, the
spectral curve, and the degrees of freedom integrated over in the reduction
process carry definite physical meaning in connection to gravity and its matrix
theory description. Referring to Ref.~\cite{Post:2022dfi} for a detailed
discussion, we here provide a quick glossary summarizing the most important
structures:

The coordinate $x$ locally parameterizing the spectral curve has the physical
interpretation of (complex) \emph{energy}. Considered as a Riemann surface, the
geometry of the spectral curve encodes information on the spectral density.
Specfically, the branch points defined by the equation $y^2-x=0$ know about the
spectral edge, $x=0$, and the spectral density $y(x)\sim \sqrt{x}$ in its
vicinity \footnote{This choice of spectral curve corresponds to the so-called
Airy or Witten-Kontsevich model and gives rise to a world-sheet theory called
topological gravity \cite{wittentopogravity1, wittentopogravity2}. In this
approximation, the JT universes simplify and are in one-to-one correspondence
with ribbon graphs associated to the Kontsevich matrix integral.}. 

The topological strings defined in the six-dimensional CY target space geometry
come in two variants, closed and open. \emph{Open strings} end on geometric
structures known as branes, and we here need to distinguish between a large set
of $D$ compact \emph{color branes} and an $\mathcal{O}(1)$ set of non-compact
\emph{flavor branes}. The data comprising the $\mathcal{O}(D\times D)$ open
string endings on color branes $i,j=1,\dots,D$ defines a `matrix', $H_{ij}$ and
the integration over it the equivalent of an ensemble average
\cite{Dijkgraaf:2002vw, Dijkgraaf:2002fc}. This structure naturally explains why
gravity obtained by dimensional reduction resembles ensemble averaged theories.
By contrast, flavor branes act as probes into the theory. As subsets of the CY,
they are defined by fixing coordinates, $x,z,\dots$, on the spectral curve and
setting either $u$  or $v$  
to zero to define non-compact brane or anti-brane
submanifolds of the CY parameterized by $(x,v)$ or $(z,u)$. Within the field
theory framework, these objects are then represented by vertex operators
$\psi(x)\equiv \exp(\Phi(x))$, and $\psi^\dagger(z)\equiv\exp(-\Phi(z))$,
respectively \cite{Aganagic:2003db, Aganagic:2003qj}. These operators probe the
information otherwise obtained by the insertion of matrix determinants
$\det(x-H)$ or $\det(z-H)$ into the numerator or denominator of an ensemble
averaged correlation function. Finally, \emph{closed strings} are interpreted as
circular mini-universes, and in the field theory are represented as primaries
$\partial \Phi(x)$. The cubic interaction of the boson with coupling constant
$\lambda_\textrm{KS}\sim \exp(-S_0)$ describes the dynamical creation and
annihilation of mini-universes in the gravitational context. 

Turning to a more concrete level, the discussion above indicates that the
state geometry correlation function Eq.~\eqref{eq:det_ratio}  affords a
representation   
\begin{align}
    &Z\equiv \Big\langle \Big\{ \psi^\dagger(z) \psi^\dagger(-z) \psi(x_1)\cdots\psi(x_4) \Big\}\Big\rangle_\mathsf{KS} 
    \label{eq:Kontsevich3}
\end{align}
in terms of brane- and anti-brane insertions into KS field theory, where normal
ordering $\{\dots\}$ removes divergent terms originating in  the OPE of the
chiral boson $\Phi(x)\Phi(x')\sim \log(x-x') + \mathrm{reg.}$ Note that the
geometric set-up for studying properties of states is different from the one
utilized to calculate spectral quantities in Ref.~\cite{Altland:2022xqx}, as
Eq.~\eqref{eq:Kontsevich3} involves an asymmetric configuration of branes and
anti-branes. Furthermore, the flavor branes have finite separation which signals
that the quantum state statistics is not accessible through a perturbative
closed string expansion, but requires knowledge of the open string sector. In
the following we outline how this correlation function in KS field theory
quantitatively agrees with that previously discussed in RMT, including for brane
coordinates $z$ in the immediate vicinity of the edge, i.e. $\Sigma$'s branch
point, $z=0$. 

We first note that the bookkeeping of all Wick contractions required by the
normal ordering of exponentiated fields generates a super Vandermonde
determinant $s\Delta(X)$ of the brane coordinates $X=(z,-z|x_1,x_2,x_3,x_4)$, which already
featured in our previous analysis, Eq.~\eqref{eq:Z_radial_integrals}:
\begin{align}
Z = \frac{1}{s\Delta(X)} \Big\langle \psi^\dagger(z)\psi^\dagger(-z) \psi(x_1)\cdots\psi(x_4) \Big\rangle_\mathsf{KS}~,
\label{eq:flavormatrix}
\end{align}
now without normal ordering. To proceed, we apply a Fourier transform to pass from the energy-like spectral
curve coordinates $(z,x_i)$ to time-like variables $(u,w_i)$. The transformed
version of the correlation
reads as 
\begin{widetext}
\begin{align} \label{app_eq:formulaZ}
Z=\frac{1}{s\Delta(X)}\prod_{i=1}^4 \int_{\mathcal{C}_0} dw_i \int_{\mathcal{C}_+}du_1 \int_{\mathcal{C}_-}du_2\,  
e^{\frac{1}{\lambda_{\mathsf{KS}}}\left(z(u_1-u_2) + \sum_{i=1}^4 w_i x_i\right)}\Big \langle
     \hat{\psi}^\dagger(u_1) \hat{\psi}^\dagger(u_2)\hat{\psi}(-w_1)\cdots\hat{\psi}(-w_4)
 \Big\rangle_\mathsf{KS}~,
\end{align}
\end{widetext}
where the  integration contours $\mathcal{C}_0,\mathcal{C}_{\pm}$ are shown
in Fig. 1 of the
supplemental material. The architecture of this representation indicates that
the diagonal matrix of  time-like variables $W\equiv \textrm{diag}(u_1,u_2|-w_1,-w_2,-w_3,-w_4)$
assumes a role analogous to that of the radial variables in the Kontsevich
framework. To corroborate this analogy, we reintroduce  normal ordering by
multiplying by the super Vandermonde $s\Delta(W)$.  What then remains to
evaluate is 
\begin{align*} 
\hat{Z}\equiv \Big \langle\Big\{ \hat{\psi}^\dagger(u_1) \hat{\psi}^\dagger(u_2)
 \hat{\psi}(-w_1)\cdots\hat{\psi}(-w_4)\Big\}\Big\rangle_\mathsf{KS}~.
\end{align*}
Thanks to normal ordering, the vertex operators can be combined into
a single exponential:
\begin{align*}
\hat{Z}&=\Big \langle\Big\{\exp\Big[-\sum\limits_{i} \hat{\Phi}(u_i)+ 
\sum_{j}\hat{\Phi}(-w_j) \Big] \Big\}\Big\rangle_\mathsf{KS}\simeq\\
&\exp\bigg(-\frac{1}{\lambda_{\mathsf{KS}}}\Big( \sum_i\Big\langle \widehat{\Phi}(u_i)\Big\rangle_0- 
\sum_j\Big\langle \widehat{\Phi}(-w_j)\Big\rangle_0\Big)\bigg)~,
\end{align*}
where the replacement $\langle \dots
\rangle_\mathsf{KS}\to \langle \dots
\rangle_0$ indicates that to leading order in $\lambda_{\mathsf{KS}}\sim \exp(-S_0)$, the
field expectation values are computed in the free theory. This final
computation, detailed in Ref.~\cite{Altland:2022xqx},
yields  $\langle
\widehat{\Phi}(w)\rangle_0 = -\int^{w} x(w')dw'$, in terms of solutions $x(w)$
of the spectral curve equation $H(x,w)=0$. Considering
the near-edge $H(x,w)=w^2-x$, and doing the integral, we obtain
\begin{align*}
    \hat{Z}=\exp\bigg(\frac{1}{3 \lambda_{\mathsf{KS}}}\Big(\sum_i u_i^3+\sum_j w_j^3\Big)\bigg)~,
\end{align*} 
Substituting this result into Eq.~\eqref{app_eq:formulaZ} we obtain an
expression equal to Eq.~\eqref{eq:Z_radial_integrals}, with the identification
of coupling constants $c=\lambda_{\mathsf{KS}}^{-1}=\exp(S_0)$ and the size of the flavor matrix determined by the number
of (anti-)branes. In this way, we establish the non-perturbative equivalence
between the theory of topological strings and the matrix integral framework in
the description of (gravitational) wave functions near the edge.

\section{Supplemental Material to ``The universality class of the first levels''}

We here  provide details for the calculation of the FS
distribution presented in the main text.

Starting from, $H = H_0 + \lambda\, H_\lambda$, with $H$ and
$H_\lambda$ drawn from the Gaussian distribution,
$P(H) dH \propto e^{-
\frac{D}{2} {\rm tr} H^2} \prod_{i, j} dH_{ij}$,
we follow
Ref.~\cite{Penner:2020cxk} to first take the Fourier transform of
Eq.~\eqref{eq:FS_distr_def} and then average over $H_\lambda$, to obtain
the 
characteristic function 
\begin{align}\label{app_eq:intermediate_FT_representation}
P_E(\omega) = \left\langle \sum_{n} \delta(E - E_n) \prod_{m\neq n} \frac{(E_m - E_n)^2}{(E_m - E_n)^2 - i \omega / D}\right\rangle_{H_0}.
\end{align}
Using  the joint-eigenvalue distribution for $H_0$, $p_D(E_1, ...,
E_D) \propto \prod_{n<m}(E_n - E_m)^\beta e^{-
\frac{D}{2}\sum_n E_n^2}$~\cite{haake1991quantum}, with $\beta = 0, 2$
for cases (i), and  (ii, iii), respectively, this becomes a determinant ratio 
\begin{align}
\label{app_eq:characteristic_form}
P_E(\omega) 
&= 
\Bigg\langle \frac{\det^{2 + \beta}(H - E)}{\prod_{j}^2 \det(H - E + a_j)}  \Bigg\rangle_{H},
\end{align}
where $a_j = \pm e^{i\pi{\rm sgn}(\omega)/4}\sqrt{|\omega|/D}$ and $H$ now is a Gaussian
distributed  Hamiltonian of size $D-1$  (this size reduction is irrelevant for $D\gg1$). 
We continue by changing variables $\omega \sim z^2$, and introducing  
placeholder energy variables $x_i$ in the determinants in the numerator which
will be taken to zero eventually. This recovers
Eq.~\eqref{eq:det_ratio}, with the distribution $P_E(g)$
given in terms of a Gaussian transform. In what follows, we consider 
 (ii, iii) with $\beta = 2$.

As a ratio of determinants, the generating function in Eq.~\eqref{eq:det_ratio}
of the main text affords a $\sigma$-model representation whose construction
follows a textbook protocol~\cite{Efetbook,Altland2023}:
\begin{align}
\label{app_eq:general_FT}
Z(X) 
\propto 
\int DA\, e^{- \frac{D}{2}\, {\rm str} A^2 - D\, {\rm str}\ln(A + X - E - is0)},
\end{align}
where $X=\mathrm{diag}(z,-z|x_1,x_2,x_3,x_4)$ carries the $z$-dependency, and
$s$ are infinitesimal imaginary increments in the boson-boson sector
guaranteeing convergence, with $s=+1$ and $-1$ for advanced, respectively,
retarded components. (Intuitively, the `${\rm str}\ln$' is the super-determinant obtained
by integrating over a Gaussian integral representation of the original matrix
determinants, and $A$ a Hubbard-Stratonovich field introduced to decouple
quartic non-linearities resulting from the $H$-average.) 

We exploit the symmetry of Eq.~\eqref{app_eq:general_FT} by introducing radial
coordinates $A = RWR^{-1}$, with $W = {\rm diag}(u_1, u_2 | -w_1, -w_2, -w_3, -w_4)$.
In this representation, the integral over the rotation matrices can be carried
out via the Itzykson-Zuber formula~\cite{superitzyksonzuber}, and  we arrive at
Eq.~\eqref{eq:Z_radial_integrals}, with the action $S(W) = - \frac{D}{2}\, {\rm
str} W^2 - D\, {\rm str}\ln (W + X - E - is0)$. 
While everything was
exact to this point, we next have to do approximations which depend on the
energy range of the tested eigenstates.

\subsection{Bulk states}

The analysis of bulk state geometries  follows the strategy of
Ref.~\cite{Penner:2020cxk}. A stationary phase analysis
stabilized by the large parameter $D\rho_E\gg1$ identifies  the saddles
$W=\frac{E}{2} \pm i\rho_E$, with the Wigner semicircle density of
states $D\rho_E = D\sqrt{1 - E^2/4}$, and the signs in the bosonic sector
fixed by the imaginary increments of $z$ for reasons of convergence.
In the four dimensional fermionic sector only the six choices of
`signature zero' (two plus and two minus signs) contribute to leading order in
$1/(\rho_E D)$. Summing over these six permutations  $s_{\sigma(i)}$ of the
standard saddle $s_0=(-1,+1|s_1,s_2,s_3,s_4)=(-1,+1|-1,+1,-1,+1)$ we arrive at
 \begin{align*}
Z(X) 
\propto 
\sum_{\sigma } &\frac{\prod_{i = 2, 4}(x_{\sigma(i)} - z)
\prod_{i = 1, 3}(x_{\sigma(i)} + z)}{z\prod_{i = 1, 3}
\prod_{j = 2, 4}(x_{\sigma(j)} - x_{\sigma(i)})}\times \nonumber\\ 
&\qquad\times e^{D {\rm str} W_0 X_{\sigma}}.
\end{align*}
Taking the limits $x_i
\to 0$, we  arrive at the characteristic function
\begin{align*}
Z(z) 
&\propto 
\left(
1 - 2i\rho_E D z 
- 
\frac{4}{3}\rho_E^2 D^2 z^2 
+ 
\frac{i}{3} \rho_E^3 D^3 z^3 \right)e^{ 2i \rho_E D z},
\end{align*}
and upon Gaussian transformation get $P_{\rm bulk}(y)=p_{\rm bulk}(y) e^{-1/y}$,
with $y = g / (D\rho_E^2)$ and the polynomial $p_{\rm bulk}$
stated in the main text. This result is exact up to corrections in 
$1/D$ and for $E = 0$ reduces to the one obtained previously in
Refs.~\cite{Sierant2019,Penner:2020cxk}.

For energies $E \searrow - E_0$ (with $-E_0$ being the energy of the lower edge),
but still outside the
support of the UFL, $E + E_0\equiv \epsilon\gg \Delta_\textrm{e}$,
the Wigner semicircle approximates to $\rho_{\epsilon} \simeq
\sqrt{\epsilon}$, defining the new scaling variable $y=g/D \epsilon$ and the
corresponding near edge bulk distribution $p_\textrm{bulk}(y)$. We note that the
same distribution would be obtained by stationary phase integration over the
Kontsevich action Eq.\eqref{eq:Kontsevich}. This follows from the fact that the
latter is but the near edge expansion of the action in
Eq.~\eqref{app_eq:general_FT} to cubic order $A$ for energies $E\approx -E_0$.~\cite{SonnerAltland21}

\subsection{Edge states}

\begin{figure}[t]
\centering
\includegraphics[width=0.4\linewidth]{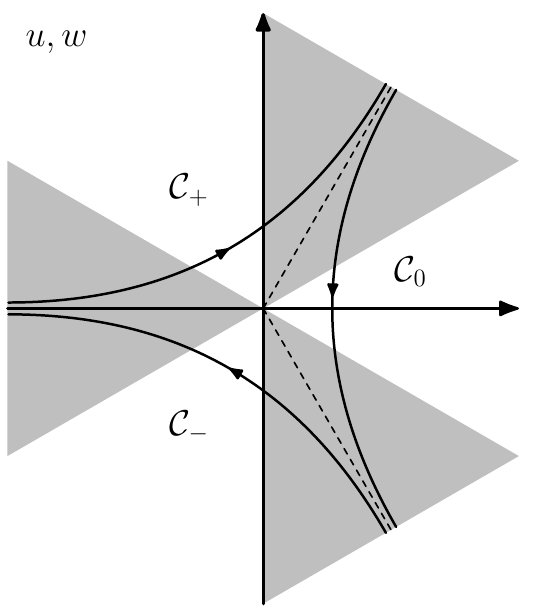}
\caption{Convergence requires the integration contours of the `Airy-integrals'
over radial coordinates to pass through the areas indicated in gray shading. 
$\mathcal{C}_0$ defines the `standard' Airy function, and
contours $\mathcal{C}_{-}$, $\mathcal{C}_{+}$ 
linear combinations of Airy and Bairy functions.}
\label{app_fig:integration_contours_Kontsevich_matrix_integrals}
\end{figure}

Now turning to the immediate vicinity of the  edge,  $\epsilon
\simeq \Delta_\textrm{e}$, stationary
phase no longer is an option. Instead, we turn to an exact integration over
$A=R W R^{-1}$ with integration contours for
the radial variables $W$ depicted in
Fig.~\eqref{app_fig:integration_contours_Kontsevich_matrix_integrals}. Broadly
speaking the fermionic radial coordinates are integrated over a contour $\mathcal{C}_0$ aligned
parallel to the imaginary axis, while the integration over bosonic variables is
over contours $\mathcal{C}_\pm$ shifted into the complex plane according to the
imaginary offset of the variables $z$. For a detailed discussion we refer to the
orignal references~\cite{SonnerAltland21,Altland:2022xqx}. 

Once again integrating over rotation matrices via Izykson-Zuber, and taking the
limit $x_i\to 0$    we arrive at
\begin{align}
Z(z) \propto z^7 \int DW\, {\rm s}\Delta(W) \Delta(W_{\rm ff}) e^{S(W)},
\end{align}
where $dW=du_1du_2dw_1dw_2dw_3dw_4$, $S(W)$ is the Kontsevich action, and taking
the limit $x_i\to 0$ has generated the additional Vandermonde determinant of the
fermionic variables $W_\textrm{ff}$. The Vandermonde determinants and the action
$S(W)$ contain the six variables $W$ in a highly symmetric fashion,
enabling us to do the integral in closed form:
\begin{align}\label{app_eq:Kontsevich_characteristic_function}
Z(z)
\propto &
q_1(\zeta) {\rm Ai}(\bar{\zeta}) {\rm Ai}(-\zeta) 
+
q_2(\zeta) {\rm Ai}(\bar{\zeta}) {\rm Ai}'(-\zeta)
+\nonumber \\ 
&+q_3(\zeta) {\rm Ai}'(\bar{\zeta}) {\rm Ai}(-\zeta) 
+
q_4(\zeta) {\rm Ai}'(\bar{\zeta}) {\rm Ai}'(-\zeta),
\end{align}
with $\zeta=j D^{2/3}z$, $\bar{\zeta}\equiv j^{-1} D^{2/3}z$, $j=e^{-2 \pi i / 3}$ and
the polynomials
$q_1(x)\simeq 6.46 - 5.91\, jx^2$,
$q_2(x)\simeq 8.86\, j + 6.46\, x - 2.03\, j^2 x^2 - 0.09\, j x^3$,
$q_3(x) = q_2^*(-x)$, and 
$q_4(x)\simeq 12.16\, - 0.90\, j x^2$. 
We finally compute the Gaussian transform over $z$ via saddle point integration
to obtain our main result \eqref{eq:result_edge}
with 
$p(x)   
\simeq 
x^{-5/2}
+
7.12\,x^{-7/2}
+
11.61\,x^{-9/2}
+
8.72\,x^{-11/2}
+
3.56\,x^{-13/2}
+
0.79\,x^{-15/2}
+
0.08\,x^{-17/2}$, 
which is exact to leading order in $1/D^{1/3}$.

\end{document}